\documentclass[10pt,twocolumn,letterpaper]{article}

\usepackage{cvpr}
\usepackage{times}
\usepackage{epsfig}
\usepackage{graphicx}
\usepackage{amsmath}
\usepackage{amssymb}
\usepackage{bbm}
\usepackage{subcaption}
\usepackage{multirow}
\usepackage{adjustbox}

\usepackage[pagebackref=true,breaklinks=true,letterpaper=true,colorlinks,bookmarks=false]{hyperref}

\cvprfinalcopy 

\def\cvprPaperID{****} 

\begin{document}
\teaser{
\centering
\begin{tabular}{c cc}
\includegraphics[width=0.32\textwidth]{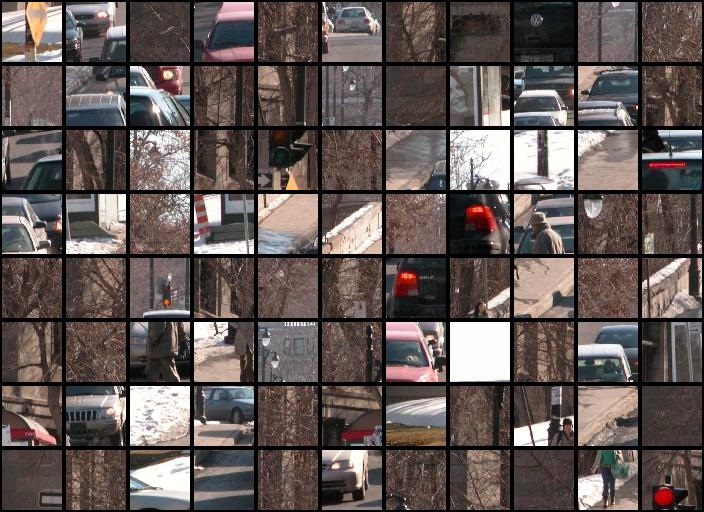} &
\includegraphics[width=0.32\textwidth]{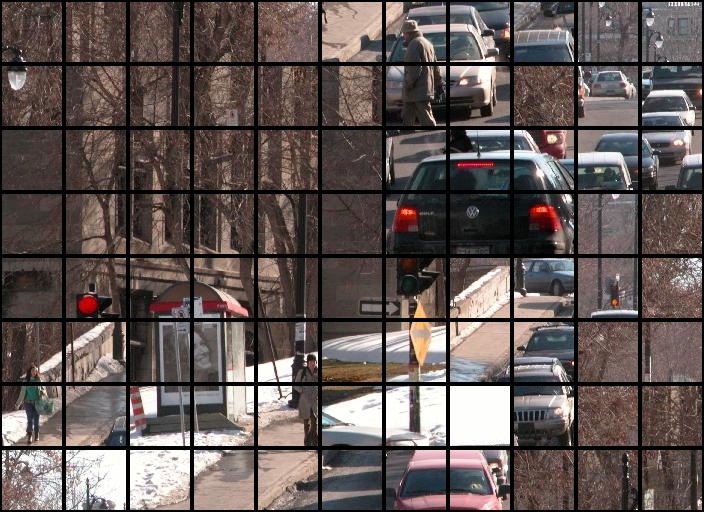} &
\includegraphics[width=0.32\textwidth]{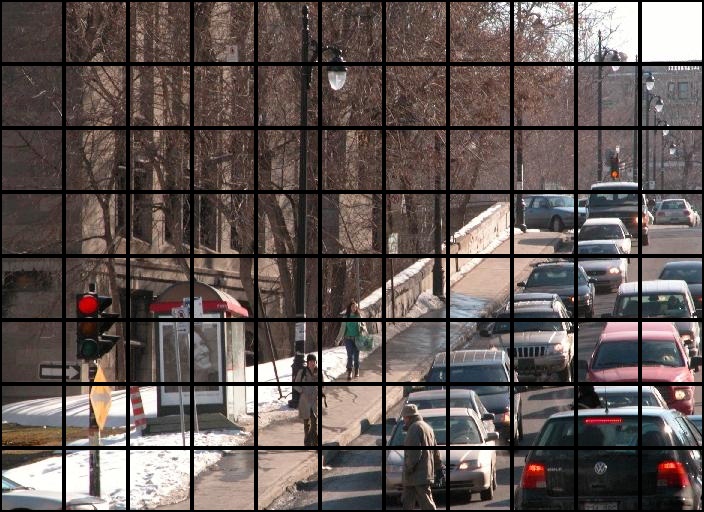}\\
input & \cite{genady}'s result & Our result\\
\end{tabular}
\caption{{\bf Puzzle solving with eroded boundaries.}
The black pixels around each puzzle piece represent missing data.
None of the puzzle piece neighbors have a shared border,
nevertheless our method is able to place the pieces.
}
\label{fig:teaser}
}
\title{Solving Jigsaw Puzzles with Eroded Boundaries}
\author{
Dov Bridger\\
Technion, Israel\\
{\tt\small db@campus.technion.ac.il}
\and
Dov Danon\\
TAU, Israel\\
{\tt\small dovdanon@post.tau.ac.il‬}
\and
Ayellet Tal\\
Technion, Israel\\
{\tt\small ayellet@ee.technion.ac.il}
}

\maketitle
\begin{abstract}
Jigsaw puzzle solving is an intriguing problem which has been explored in computer vision for decades.
This paper focuses on a specific variant of the problem---solving puzzles with eroded boundaries.
Such erosion makes the problem extremely difficult, since most existing solvers utilize  solely the information at the boundaries.
Nevertheless, this variant is important since erosion and missing data often occur at the boundaries.
The key idea of our proposed approach is to inpaint the eroded boundaries between puzzle pieces and later leverage the quality of the inpainted area to classify a pair of pieces as "neighbors or not".
An interesting feature of our architecture is that the same GAN discriminator is used for both inpainting and classification; training of the second task is simply a continuation of the training of the first, beginning from the point it left off.
We show that our approach outperforms other SOTA methods.
\end{abstract}
\maketitle
\section{Introduction}
\label{sec:Introduction}
Jigsaw puzzle solving is important in many applications, including image editing \cite{3patch_transform}, biology~\cite{13genomic_puzzle}, archaeology~\cite{1wall_paintings,2auto_assembly_from_multiple_photos,8computer-aided_reconstruction} and recovering shredded documents or photographs~\cite{2auto_assembly_from_multiple_photos,25shredded_documents,26strip-shredded} to name a few.  
The problem was proven to be NP-complete~\cite{Demaine}.
Nevertheless, algorithms have been proposed to solve various types of puzzles.

This paper focuses on the case where we are provided with an unordered set of non-overlapping square image fragments.
We aim to find the correct positioning to reconstruct the image.
A variety of solutions have been proposed~\cite{5probabilistic_jigsaw,Alpher28,genady,17fully_automated,Alpher29,Alpher30,Alpher26}, achieving excellent results.
This success has led to attempts to cope with more challenging cases, in which the basic problem is relaxed.
For instance,~\cite{genady} looked at the case of missing pieces and mixed puzzles, whereas~\cite{Gur_2017_ICCV} allowed the pieces to be "brick"-like.
Puzzles with eroded boundaries were recently handled in~\cite{shortest_path}.
Their method works for $3 \times 3$ puzzles and presents good results for this case.
It cannot be extended to handle larger puzzles dues to its exponential complexity and its reliance on learning absolute positions of pieces.

We also look at the latter variant.
The goal is to reconstruct a 2D image from square pieces of a photo, where the edges of the pieces are damaged, as demonstrated in Figure~\ref{fig:teaser}.
Thus, two adjacent pieces do not have a continuous border between them.
Eroded boundaries may be found in a variety of real-world applications, such as in restoration of old documents and in archaeology. 
Our solution is general and scalable.

Previous methods for solving large jigsaw puzzles cannot be utilized, since they rely only on information from the boundaries to measure the compatibility of a pair of pieces. 
In fact, none of the above-mentioned methods use the color information in any of the pixels that are more than $2$ pixels away from the boundary of a piece.  
Thus, with only a single row/column of pixels removed from the borders of each part, these methods might fail. 

\begin{figure*}[htb]
\centering
\includegraphics[width=0.7\paperwidth]{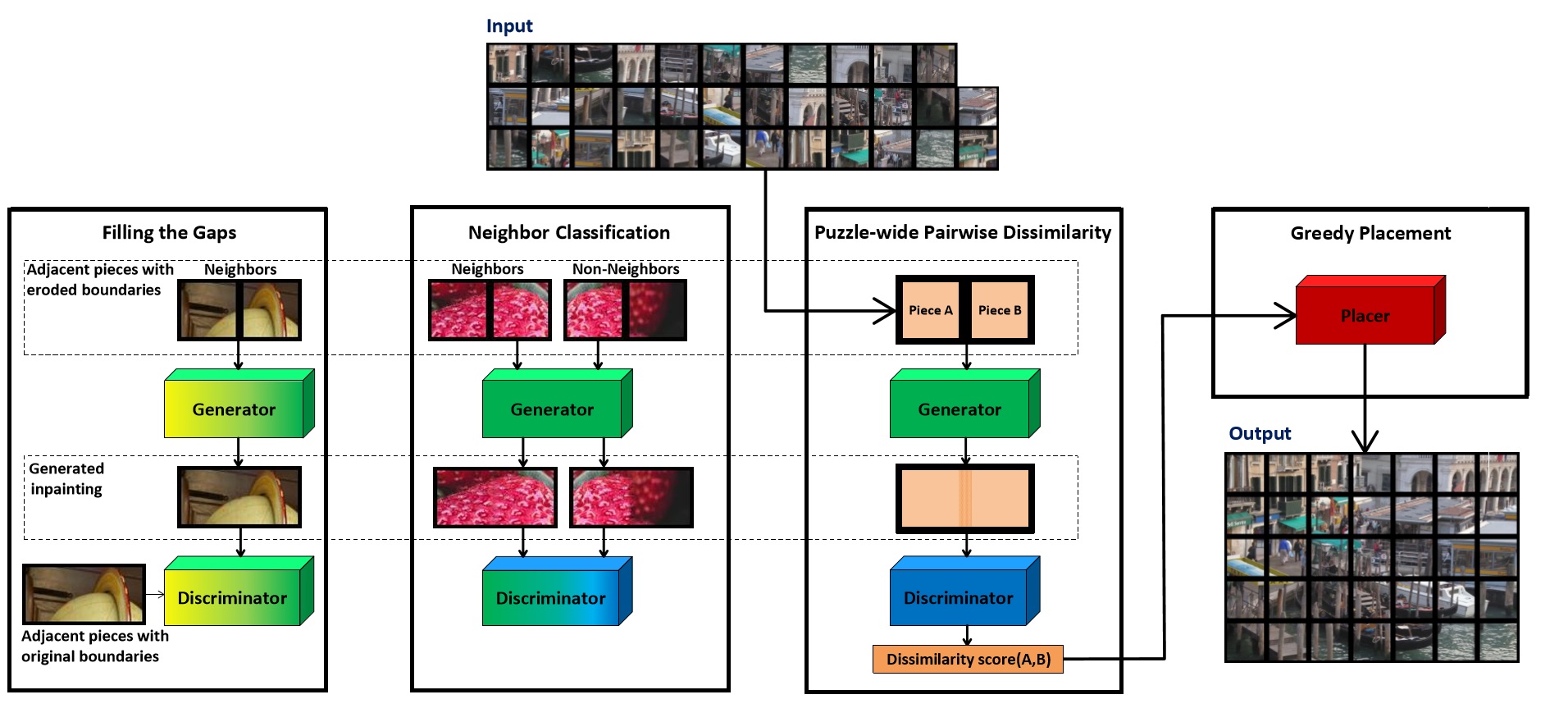}
\caption{
{\bf Model architecture.}
Our model consists of four stages: gap filling, neighbor classification, pairwise dissimilarity, and placement. 
The two blocks to the left are used only for training, whereas puzzle solving starts at the third block.
The change of color in the network (from yellow to green, or from green to blue) represents training, whereas a constant color (green or blue) represents inference only.
The generator and discriminator networks carry on from block to block. 
The dashed lines indicate that all the images within it are of the same type, as explained in the text on the left.
}
\label{fig:basic_model_architecture}
\end{figure*}

Our method is based on the following key ideas.
First, for each pair of puzzle pieces we use the known pixels in order to generate the missing pixels in the gap, along their joining borders.
This is done by a custom-fitted and trained GAN-based inpainting algorithm. 
Second, our core assumption is that our inpainting results between neighboring pieces will be recognizably better than those between non-neighbors.
This allows us to train a neighbor classifier that assesses the inpainting quality between any two pieces and compute the probability that they are neighbors.
Our novel pairwise dissimilarity metric is inversely related to that probability. 
Finally, like most of previous works, we use a
greedy placement method that is based on our pairwise dissimilarity scores to solve the puzzle.

A compelling component of our model is that the learning from the inpainting task is transferred to the neighbor/not-neighbor classification task, using an identical architecture.
We show that the weights that were learned in the former significantly boost the results of the latter.
We demonstrate state-of-the-art results for commonly-used datasets~\cite{5probabilistic_jigsaw,17fully_automated}.

Our contribution is hence threefold.
First, we present a novel and efficient puzzle solver for the case where the pieces have eroded boundaries.
Second, the key idea is that when information is missing, we can generate it; then, based on the quality of the generated information, we can draw conclusions regarding 
similarity.
This idea of using the discriminator at inference time can be used in other applications, for instance in image retargeting or face generation to further improve the generated images.
Last, but not least, we present state-of-the-art results for the case of puzzles with abraded boundaries.

\section{Related Work}
\label{sec:related}
The jigsaw puzzle problem has existed since long before it was first solved by a computer.
The first computational jigsaw solver was proposed in 1964  and was able to handle nine-piece problems~\cite{Alpher01}.
In~\cite{Demaine} it was proved that the problem is NP-complete.
Therefore, it is impossible to solve the problem accurately for puzzles of non-trivial size.

\vspace{-0.1in}
\paragraph{Square-piece puzzle solving.}
Most works in computer vision assume that the input consists of equal-size square  pieces of an image. 
The first solver was introduced by~\cite{5probabilistic_jigsaw}, where a greedy algorithm and a benchmark were proposed.
The algorithm discussed in~\cite{Alpher26} improved the results by using a particle filter.
In~\cite{17fully_automated} the first fully-automatic solver was introduced.
It was based on a greedy placer and on a novel prediction-based dissimilarity.
The method was generalized by~\cite{Alpher28} to handle parts of unknown orientation.
A considerable improvement for the case of unknown orientation was demonstrated in~\cite{Alpher30}, by adding "loop constraints" to~\cite{Alpher28}.
Rather than pursuing a greedy solver, \cite{Alpher29} presented a genetic algorithm that was able to solve large puzzles.
In~\cite{genady}, the compatibility function takes advantage of both the similarity between the pieces and the reliability of this similarity. 
Furthermore, during placement, the piece that minimizes the likelihood of erring  is selected, regardless of its location.

Variants of the basic problem were investigated.
In particular, the method of~\cite{genady} handles puzzles with missing pieces, as well as
concurrently solving multiple puzzles whose pieces are mixed together.
In~\cite{Gur_2017_ICCV} the problem is extended  to consider rectangular pieces that could be placed next to each other at arbitrary offsets along their abutting edges.

In~\cite{shortest_path}, the variant addressed is the one studied here: solving puzzles with abraded boundaries.
The proposed algorithm manages to accurately solve $39\%$ of $9$-piece puzzles.
This is done by predicting the global positioning of the image fragments with respect to a $3 \times 3$ grid.
This method is exponential and thus unscalable, as it tests all possible locations of the $9$ pieces.
Furthermore, relying on learning the position of each piece, rather than on the dissimilarity between the pieces, makes the solution suitable for datasets of a certain structure of images (e.g. images of a centered object against the background).

\vspace{-0.1in}
\paragraph{Other types of puzzles.}
Some works assume general-shaped pieces of natural images.
In~\cite{ZhangL14} a graph-based optimization framework is proposed, utilizing both the  geometry of the fragments and their color. 
In~\cite{Huang:2013:MGT:2508363.2508373}, the focus is on the gaps between the pieces.
However, the user provides an approximate initial placement and the algorithm refines the registration by simultaneously aligning all the pieces. 
In both cases, the number of pieces is rather small.

Geometry-based 2D apictorial puzzles were studied in~\cite{DBLP:conf/compgeom/GoldbergMB02}, where a "human-like method" is introduced,
identifying corners and frame pieces, and proceeding to greedy placement from the outside frame towards the interior of the puzzle. 
In~\cite{geometry_2D} the {\em Euclidean signature}~\cite{euclidean_signature} is used to match curves while relaxing the constraints of pieces having four well-defined sides and a rectangular puzzle.

Finally, finding a way to bring ruined ancient artifacts back to life has sparked the imagination of scientists for many years~\cite{willis2008computational}.
Some works have focused on matching fragments based on 3D geometric features, rather than on performing the full re-assembly~\cite{FunkhouserJOCCH11,Tolerfranklin}.
These 3D geometric features were used in~\cite{ThomasFunk:2017:WPR} for global re-assembly, based on a genetic algorithm, where a highly challenging archaeological example is solved.

\section{Model}
\label{sec:model}

We are given an unordered set of square image fragments with eroded boundaries.
We assume that 
(1) like in previous works, each fragment has the same size;
(2) the maximum extent of the damaged area is known;
(3) the orientation is unknown (optional).  
Our goal is to reconstruct the image, i.e. to find the correct position and orientation of each piece.
Our model, illustrated in Figure~\ref{fig:basic_model_architecture}, is based on the following three key ideas.

First, in order to determine how likely two parts are  to be neighbors, we fill the gap between them. 
If this filling is "realistic", they are more likely to be adjacent than otherwise.
We realize this idea in two steps.
We first learn how to perform inpainting in our special case.
This is done using a GAN setup where a generator learns how to inpaint the gap. A discriminator helps it train using examples of adjacent neighbors with and without a gap in between.
Section~\ref{sec:inpainting} provides details.
Then, we learn how to discriminate true neighbors from non-neighbors based on the inpainting quality.
We show that we can use the same generator to further train the very same discriminator (in terms of architecture and weights) in this classification task simply by feeding it with inpainted examples of neighbors and non-neighbors.
Unlike the previous step that used only true neighbors, in this step an equal representation of the much wider domain of non-neighbor examples is used.
The discriminator will learn to discriminate the inpainted gaps between neighboring pieces from those between non-neighbors. 
We expect it to be successful in doing so because the gaps between neighbors should have been filled in a plausible way, while gaps between non-neighbors should have an unrealistic inpainting result.
After completing training, the discriminator, when given an input of two puzzle pieces, will output the probability that these two pieces are neighbors in the puzzle solution.
Section~\ref{sec:classification} elaborates on this step.

The second idea is that we can use the aforementioned discriminator to compute a continuous value representing pairwise dissimilarity between pieces.
We set the dissimilarity between two pieces in a given direction to be inversely related to the probability outputted from the discriminator, which is the likelihood of adjacency.
See Section~\ref{sec:classification} for details.

Finally, given the pairwise dissimilarity scores between all pairs of pieces, we apply the greedy placement of~\cite{genady}.
Briefly,
pieces are placed in an iterative manner.
At every iteration, the placer selects the best piece from the candidate pool in order to achieve maximum compatibility between it and the existing pieces on the board.
Compatibility favors pairs having low dissimilarity with each other but also high dissimilarity with all the others.
Each placed piece adds its most compatible neighbors to the pool of candidates to be placed. 

\section{Filling the Gaps}
\label{sec:inpainting}
Given a pair of pieces, our goal is to fill the missing pixels in the gap of unknown pixels between them in a realistic way.
Later, we will identify the real neighbors based on the quality of this inpainting.

\begin{figure}[tb]
\begin{tabular}{ccc}
(a) Input&
(b) \cite{IizukaS017}&
(c) Ours\\
\includegraphics[width=0.15\textwidth]{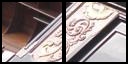}&
\includegraphics[width=0.15\textwidth]{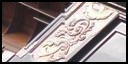}&
\includegraphics[width=0.15\textwidth]{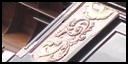}\\
\includegraphics[width=0.15\textwidth]{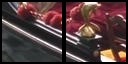}&
\includegraphics[width=0.15\textwidth]{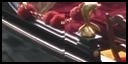}&
\includegraphics[width=0.15\textwidth]{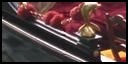}\\
\includegraphics[width=0.08\textwidth]{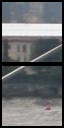}&
\includegraphics[width=0.08\textwidth]{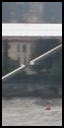}&
\includegraphics[width=0.08\textwidth]{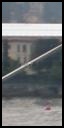}\\
\end{tabular}
\caption{
{\bf Inpainting results.}
  For local fixed masks, our inpainting method does a better job preserving structures and edges. 
  In all the examples, our inpainting continues the structures seamlessly across the inpainted boundary, whereas~\cite{IizukaS017}'s results create artifacts, breaking the frames (top two rows) or breaking the white bar (third row).
  }
\label{fig:inpainting_comparison}
\end{figure}

Various methods exist for image inpainting~\cite{image_melding,IizukaS017, yeh2017semantic, Huang:ImageCompletion14, ZhuHTXH16, BaekCK16, GilbertCJP18}. 
Most state-of-the-art methods use some form of a {\em Generative Adversarial Network (GAN)}~\cite{gan}. 
We have tried several methods~~\cite{image_melding,IizukaS017, yeh2017semantic} and found that while their results are excellent for inpainting full images, they do not work as well in our case as demonstrated in Figure~\ref{fig:inpainting_comparison}, where we compare our results to the best method, \cite{IizukaS017}.
This may be explained by the fact that our images are small (puzzle parts), and therefore a global discriminator as used in \cite{IizukaS017} has no reliable connections to learn.
Instead, we introduce a model that suits our input type---a rectangular image comprised of two small adjacent square images, representing two neighboring puzzle pieces.
We too train a GAN model, inspired by~\cite{pix2pix}.
However, we train it using a single rectangular mask, depicting the gap to be inpainted.
Hereafter, we elaborate on architecture, training and losses.

\vspace{-0.1in}
\paragraph{Architecture.}
When given an input of any two adjacent puzzle pieces, our GAN model will learn to generate a plausible solution for the gap of missing pixels across their border.
We limit the generator to change only the pixels that are in the gap. 

{\bf Generator:}
Our generator is an encoder-decoder architecture with skip connections between every symmetric pair of layers and with no bottleneck between the encoder and the decoder.

The input image enters the encoder and is reduced in spatial dimensions by factors of $2$ until it exits and transfers into the decoder.
In the decoder the process is reversed except that each layer is fed also with outputs from the corresponding layer in the encoder through skip connections.
 The number of layers is set so that at the transition from the encoder to the decoder the smallest dimension (the height) will be $1$; it is set to $6$ in our case.
The input to the generator consists of two neighboring puzzle pieces, placed side by side with a gap of missing pixels in between. 
Vertically adjacent neighbors are used as well, but they are rotated by $90^\circ$ before entering the model.

The generator generates the entire image, rewriting all of the pixels. In order not to change the pixels that we already know, the output layer is formed by copying all of the known pixels directly from the input, while the rest are generated through the network.
Based on that output, the loss, which will be later defined (Equation~\ref{eq:generator_loss}), is computed.

{\bf  Discriminator.} 
The input image is center cropped in width to half the size.
We focus the discriminator on the center of the dual-piece image 
in accordance with our key idea of using the inpainting quality of the gap to determine the likelihood that the pieces are neighbors.

We use a Markovian Discriminator with a $3$ layer encoder (similar to the generator's encoder).
The final layer contains  $6 \times 6$ patches, each with a single probability value, that are averaged to give the final probability output.
This discriminator captures high frequencies in local patches~\cite{pix2pix} which explains the better conservation of structures that we achieve. 

\vspace{-0.1in}
\paragraph{Training.}
A training example consists of two adjacent puzzle pieces cut out of some training image where the boundary pixels of each piece were removed.
The width of the boundary to be removed is a parameter that can be changed to control the level of difficulty.

We wish the generator's loss function to consider both the accuracy of the reconstructed area and the realism of the image.
Therefore, the loss function is comprised of two components, each addresses a single requirement.
Let $GI$ be the generated image,
$D(GI)$ be the discriminator's prediction for the generated image $GI$,
$OB$ be the area in the original image where the pixels were removed,
and $GB$ be the area in the generated image where pixels were generated to replace the missing pixels in the input.
Inspired by~\cite{pix2pix}, the generator loss function is defined as:
\begin{equation}
    G_{Loss} = BCE(D(GI), 1) + \lambda * L_1(OB, GB).
    \label{eq:generator_loss}
    \end{equation}
The first term, which handles realism, measures the binary cross entropy ($BCE$) between the discriminator's prediction on that output and $1$.
This is so since during training the discriminator learns discriminating real images from generated images. 
So, if we want to ask how realistic a generated image is, we can ask how close the discriminator was to classifying it as 'real'.
In the second term, $L_1(x,y)$ is the averaged $L_1$ difference between the generated pixels and the original pixels. 
In this equation, $\lambda$ is a hyper-parameter to balance the two loss components, set to $100$ in our model.

 The second loss function we define is for training the discriminator. It represents the prediction error on original images and inpainted images.
Let $OI$ be the original image of adjacent puzzle pieces before the boundary pixels were removed.
The discriminator loss function is defined as
	    \begin{equation}
    D_{Loss} = \frac{BCE(OI, 1) + BCE(GI, 0)}{2}.
    \label{eq:discriminator_loss}
    \end{equation}
The goal of $BCE(OI,1)$ is to make sure that the discriminator classifies original images as 'real'; similarly, the goal of $BCE(GI,0)$ is to classify generated images as 'fake'.

Instead of providing the discriminator with the full image, we found that better results are obtained when using a cropped version of the input, focusing on the center of image---the generated pixels and their surroundings.
This observation is in accordance with our key idea of using the inpainting quality of the gap to determine the likelihood that the pieces are neighbors.

A note on implementation:
 We trained the GAN model for $48$ epochs over $45,000$ examples using a batch size of $1$, a learning rate of $0.0002$ for the generator, and $0.0001$ for the discriminator.
 
\section{Positive/Negative Neighbor Classification}
\label{sec:classification}
After completing filling the gaps, we move on to the second training stage---learning neighbor classification.
Our goal is to classify  any given pair of puzzle pieces as a positive pair (true neighbors) or as a negative pair (not neighbors), based on the quality of the inpainting performed between them.
Moreover, we do not settle for a binary classification, but rather aim at computing a continuous score for each pair of puzzle pieces.
This is so since the score should represent the dissimilarity of the pair, which would be the basis for placement.

For this task we use the same discriminator that was used to train the inpainting network in Section~\ref{sec:inpainting}.
Furthermore, we pick up exactly where the network left off at the conclusion of the inpainting, loading the weights that were learned.
The rational behind this is that prior acquaintance with the way gaps between puzzle pieces should and should not be filled serves as an excellent initialization point for learning neighbor classification.

\begin{figure}[tb]
\centering
\includegraphics[height=0.350\textwidth]{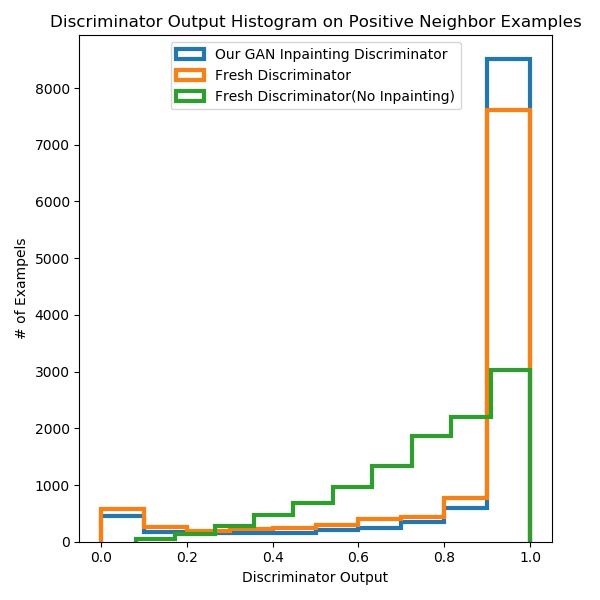}
\caption{{\bf Classification discriminator.}
Our discriminator yields an average probability output of $87.9\%$ on $11,000$ examples of positive neighbors.
A fresh discriminator yields $83.7\%$ if trained with inpainted pairs, or $75.9\%$ if the inpainting stage was skipped altogether. This shows that inpainting the gap between
pieces is critical in order to compute their neighbor probability. Moreover, using the 
same discriminator from the inpainting phase boosts results.}
\label{fig:fresh_discriminator}
\end{figure}

Figure~\ref{fig:fresh_discriminator} provides empirical evidence supporting this notion.
It shows that when given positive neighbor examples as input, the pre-trained discriminator from the inpainting phase produces higher probability outputs than a fresh discriminator trained from scratch.
Furthermore, a fresh discriminator that was trained to classify neighbors without first inpainting the gaps in the training and testing examples produces much lower outputs.

As far as negative neighbors are concerned, both our discriminator and the fresh discriminator are able to identify them very well, averaging an output of about $2\%$ for those examples. 
A fresh discriminator that was trained to classify neighbors without inpainting the gaps failed completely, averaging an output of $43.5\%$.
This proves how absolutely critical the inpainting phase in our solution is.
Since we will use these probabilities to calculate continuous dissimilarity scores, it is imperative to get the output as close as possible to $100\%$ for positive neighbors ($0\%$ for negative) and not settle for a binary (i.e. above/below $50\%$) result, which would suffice for an ordinary classification task.

\begin{figure*}[t b]
\centering
\begin{tabular}{ c c c }
(a) Ground Truth&
(b) \cite{genady}'s result&
(c) Our result\\
\hline
\includegraphics[width=0.27\textwidth]{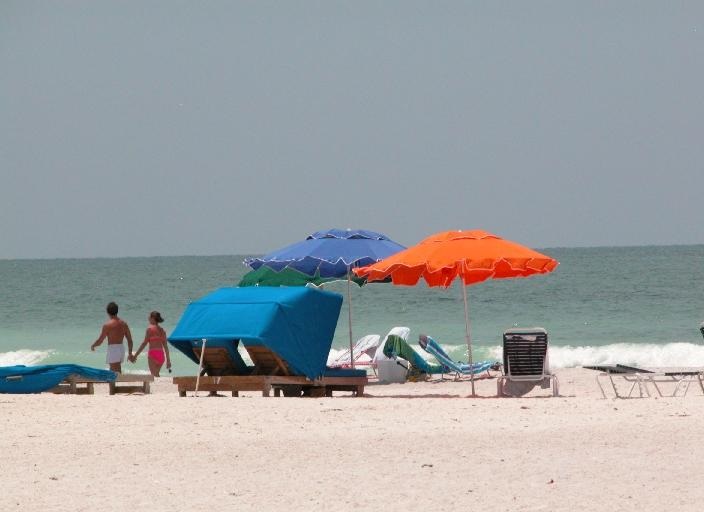}&
\includegraphics[width=0.27\textwidth]{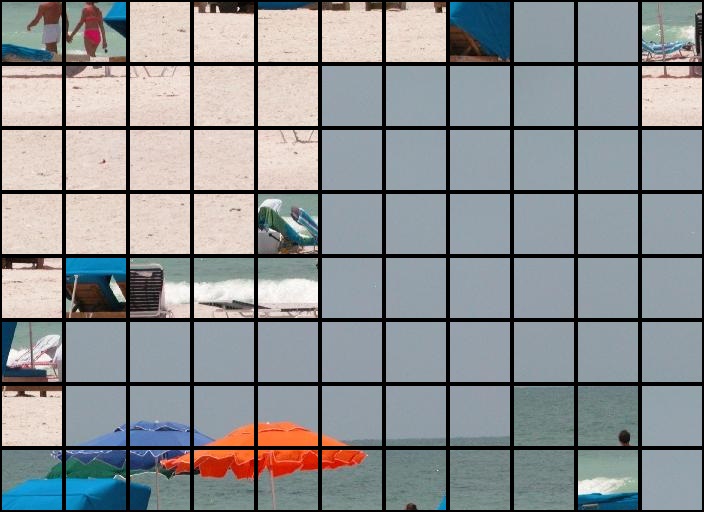}&
\includegraphics[width=0.27\textwidth]{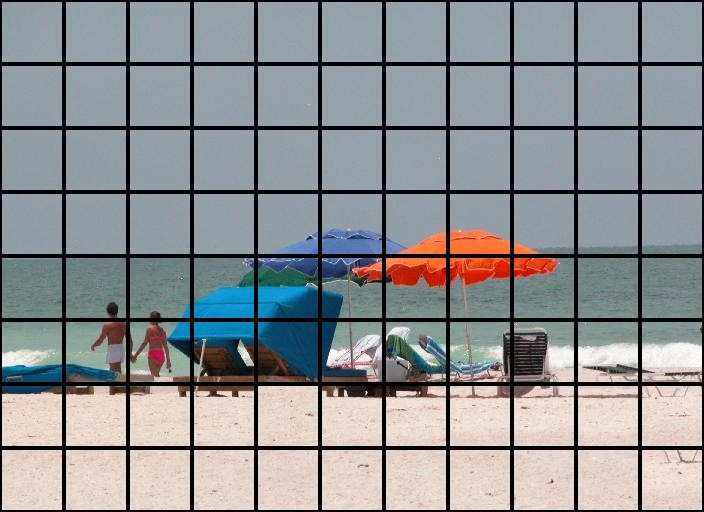}\\

\hline
\end{tabular}
\caption{
{{\bf Qualitative results}.}
Even with an erosion extent of only $7\%$ of piece size, our results exhibit much closer resemblance to the ground truth (88 pieces).}
\label{fig:qualitative_gap_4}
\end{figure*}

\vspace{-0.1in}
\paragraph{Training.}
Let $I$ be an original image of two puzzle pieces after their boundary pixels were removed and let $GI$ be the output of the pre-trained generator given image $I$ as input. 
Let $I_P$ denote that image $I$ is a pair of positive neighbors and similarly $I_N$ denote non-neighbors, such that one of the pieces is the same as in $I_P$, but the other is randomly selected from the other pieces in the puzzle.

In each training iteration we provide the discriminator with $2$ examples: $GI_P$ and $GI_N$. 
The positive example ($GI_P$) will be labeled as $1$ (true neighbors) and the negative example ($GI_N)$ as $0$. 
Note that when the discriminator was trained in the previous (inpainting) stage, examples such as $GI_P$ were labeled $0$, whereas in this stage they are labeled $1$ .
This is so because now, when the generator is trained and is able to realistically complete missing gaps between true neighbors, we want it to consider these examples as 'real' and discriminate them from 'fake' attempted completions of missing gaps between non-neighbors.
Note that even though the the number of non-neighbor pairs is much larger than that of true neighbors, the training data for the discriminator is kept balanced by providing one of each type in every iteration.

Let $D(GI)$ be the discriminator's prediction on image $GI$. 
The loss minimized in the discriminator training consists of two standard components, $BCE(D(GI\_p), 1)$ and $BCE(D(GI\_n), 0)$, where the first is in charge of detecting positive neighbors and the second is in charge of detecting negative neighbors.
It is defined as
\begin{equation}
D_{Loss2} = \frac{BCE(D(GI\_p), 1) + BCE(D(GI\_n), 0)}{2}.
\label{eq:discriminator_post_loss}
\end{equation}

Implementation-wise, we train the discriminator for $40$ epochs in this stage using a learning rate of $0.0002$. 
The training examples in this stage were not used in the previous GAN-inpainting stage because we want the discriminator to handle generated (inpainted) images that were created using unseen-before input.

\vspace{0.1in}
\noindent
{\bf Computing pairwise dissimilarity.}
After the training is complete, we use the discriminator to calculate pairwise dissimilarity between each pair of puzzle pieces in each of the four directions (up, down, right, left).
The key idea is to express how foreign two pieces are to each other across a certain boundary.
That is, the lower the prediction our discriminator makes on a certain pair to be neighbors, the higher dissimilarity value it should produce.  

Let $I_{xyd}$ be a joint image of puzzle pieces $x$ and $y$ with their boundaries removed, placed adjacent to each other in direction $d$. 
Let $G(I)$ be the generator's output on image~$I$ and $D(GI)$ be the neighbor probability prediction of the discriminator on a generated image $GI$.
The dissimilarity value between $x$ and $y$ in direction $d$ is defined as 
\[ -\log{D(G(I_{xyd}))}.\] 
In this equation, the $-\log$ is taken to convert the probability value in the range $[0, 1]$ to an inversely related dissimilarity value in the range $[0, \infty]$, while avoiding unnecessary steep changes.
Once the dissimilarity values are computed, the placement algorithm is applied to solve the puzzle.

\section{Results}
\label{sec:results}

We applied our method to solve the puzzles for all the images in three commonly-used datasets~\cite{5probabilistic_jigsaw,17fully_automated},
each containing $20$ images. 
Each image is cut into  $64 \times 64$-sized pieces, yielding puzzles containing $70$, $88$, and $150$ pieces.
We note that without gaps smaller pieces
could be used, but this is prohibitive in the in our case since the pieces will lack meaningful information.
Since no previous work exists that solves large puzzles with gaps between pieces, our results are compared to those of~\cite{genady}, which is state-of-the-art for regular square pieces, and for puzzles with missing pieces.
We adjust their solver to consider the outermost known pixels in each piece to be the actual boundary.

\begin{figure*}[h]
\begin{tabular}{ c c c c }
& 
(a) Ground Truth&
(b) \cite{genady}'s result&
(c) Our result\\
\hline
70 pieces& & & \\

Erosion = $7\%$ &\includegraphics[width=0.24\textwidth]{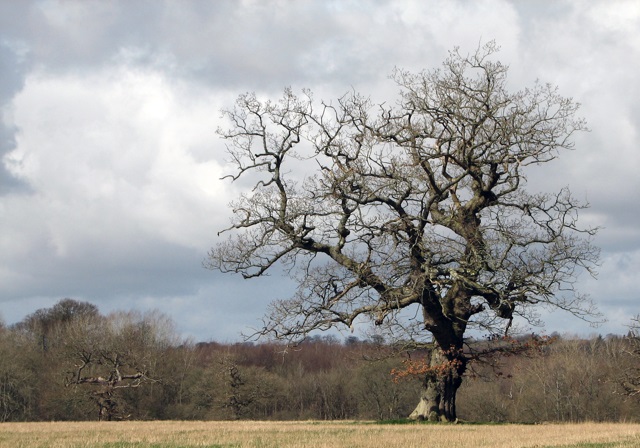}&
\includegraphics[width=0.24\textwidth]{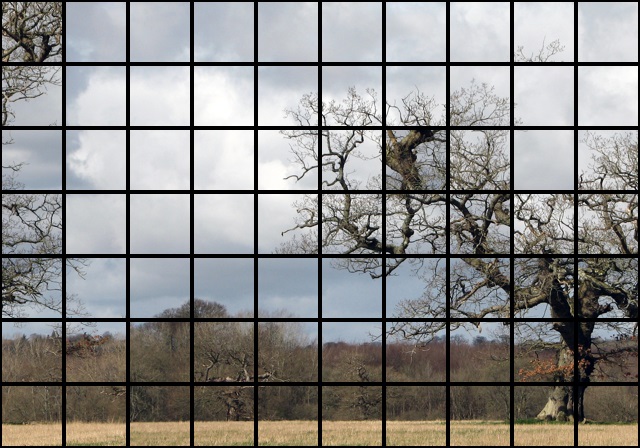}&
\includegraphics[width=0.24\textwidth]{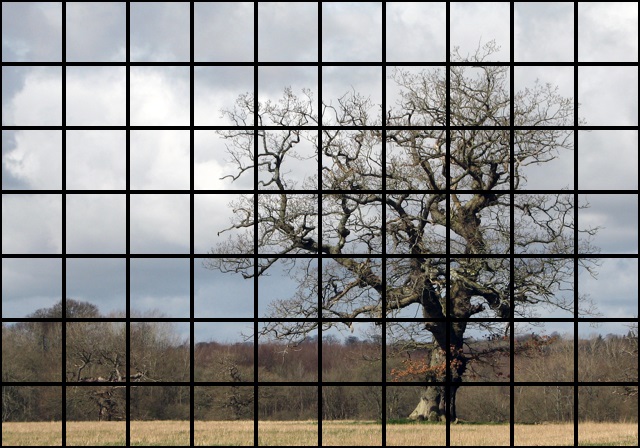}\\
\hline
Erosion = $14\%$ & &
\includegraphics[width=0.24\textwidth]{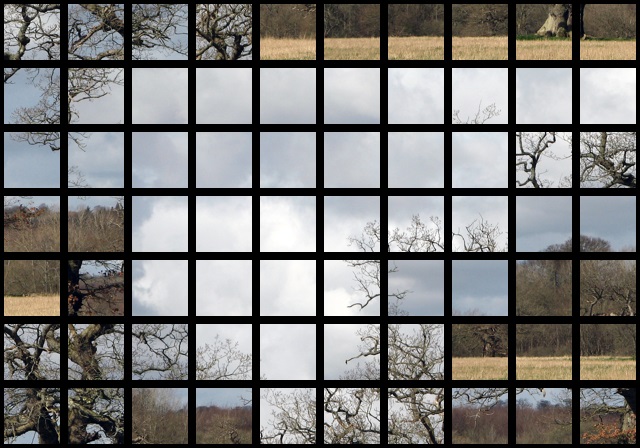}&
\includegraphics[width=0.24\textwidth]{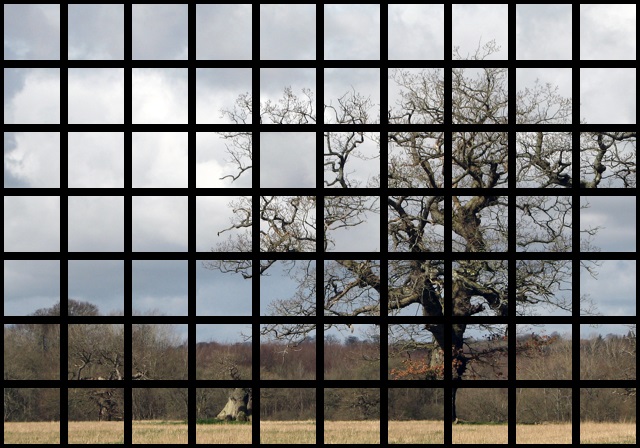}\\
\hline

88 pieces& & & \\

Erosion = $7\%$ &\includegraphics[width=0.24\textwidth]{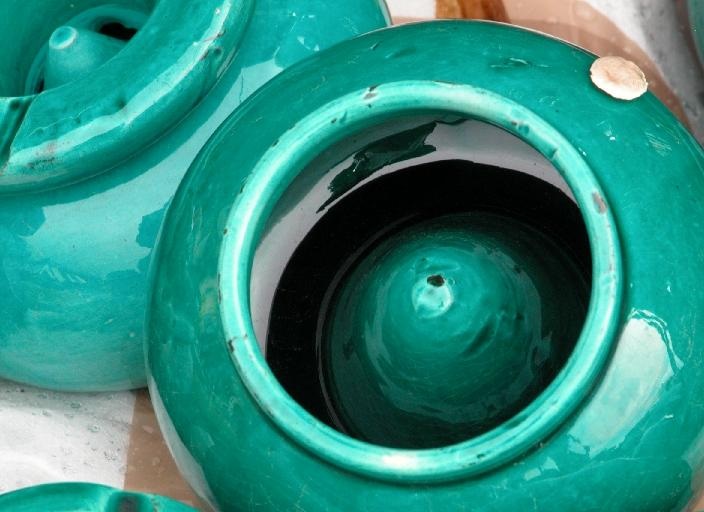}&
\includegraphics[width=0.24\textwidth]{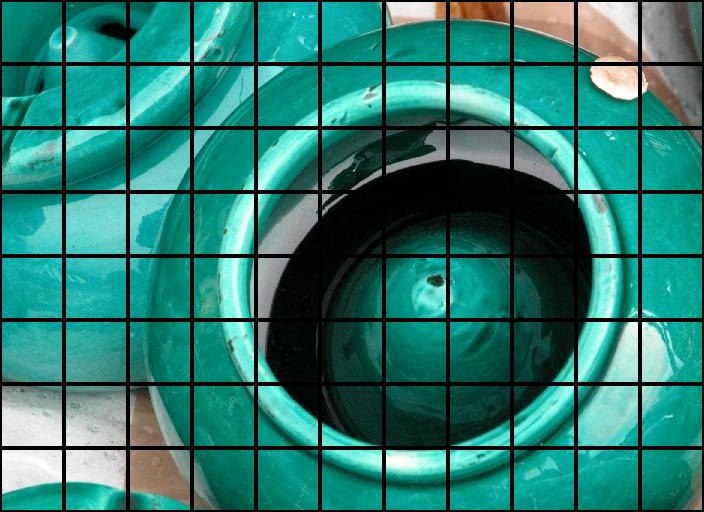}&
\includegraphics[width=0.24\textwidth]{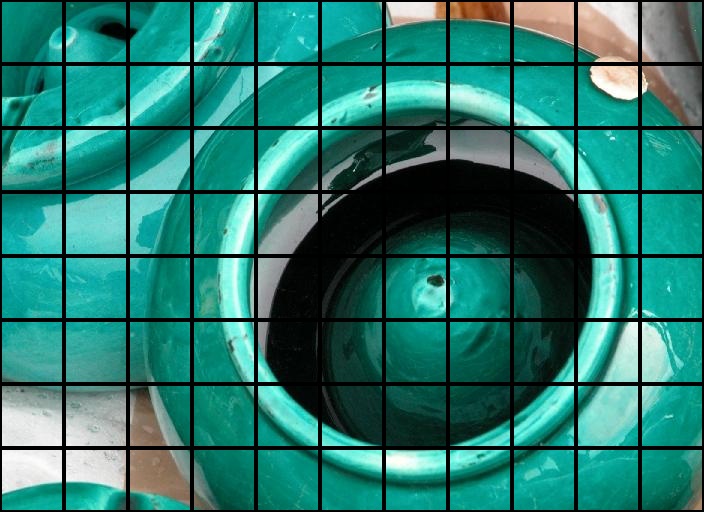}\\
\hline
Erosion = $14\%$ & &
\includegraphics[width=0.24\textwidth]{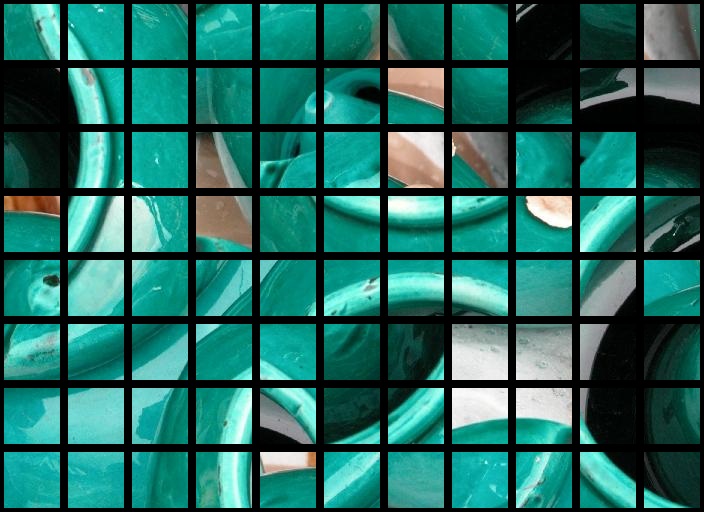}&
\includegraphics[width=0.24\textwidth]{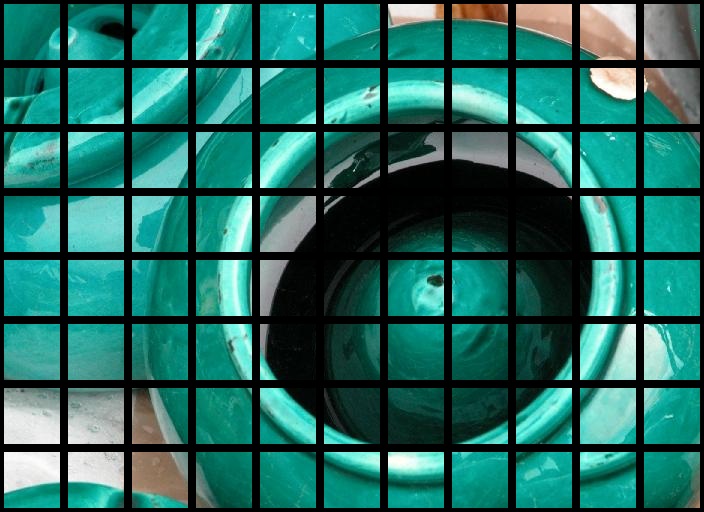}\\
\hline

\end{tabular}
\caption{
{\bf Qualitative results (erosion extent of $7\%,14\%$ of piece size).}
The larger the gaps, the worse the results. 
However, the negative effects in all examples are more dramatic in~\cite{genady}'s results.}
\label{fig:qualitative_gap_8}
\end{figure*}

\begin{figure*}[h]
\begin{tabular}{ c c c c }
& 
(a) Ground Truth&
(b) \cite{genady}'s result&
(c) Our result\\
\hline
70 pieces& & & \\

Erosion = $7\%$ &\includegraphics[width=0.24\textwidth]{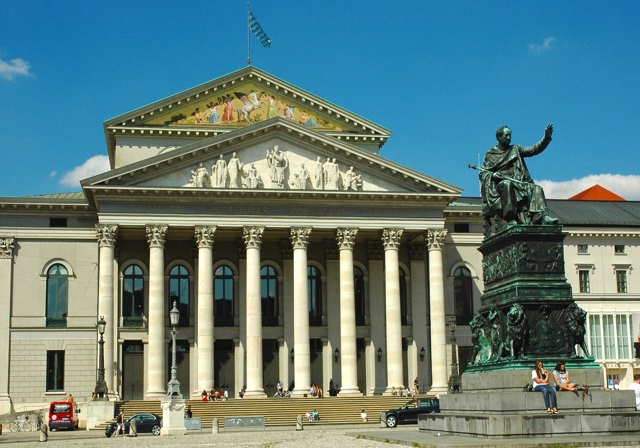}&
\includegraphics[width=0.24\textwidth]{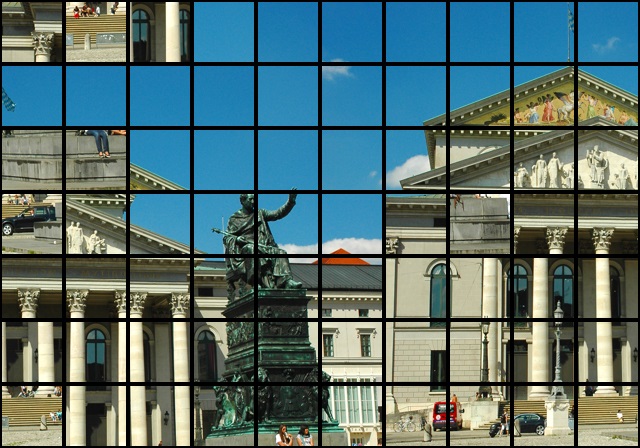}&
\includegraphics[width=0.24\textwidth]{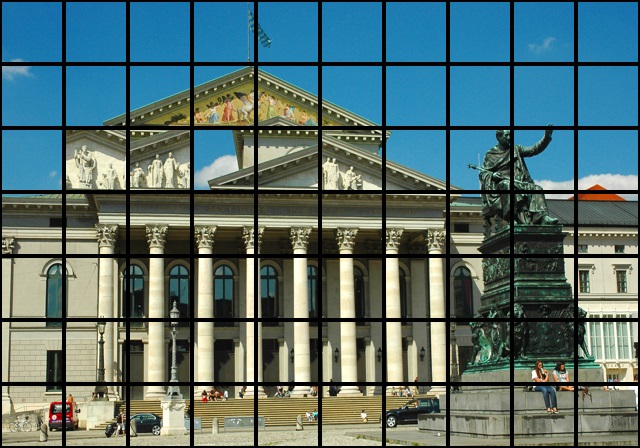}\\
\hline
Erosion = $14\%$ & &
\includegraphics[width=0.24\textwidth]{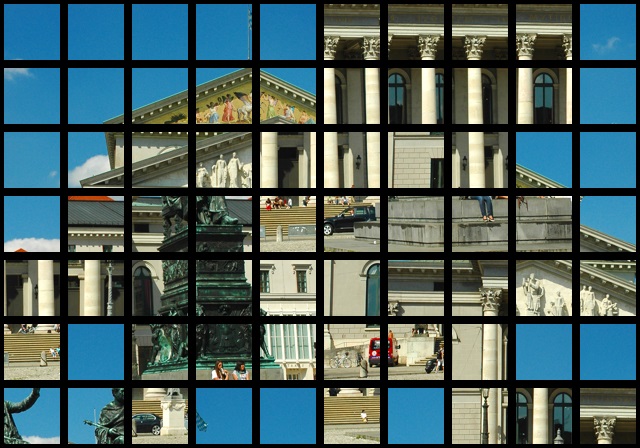}&
\includegraphics[width=0.24\textwidth]{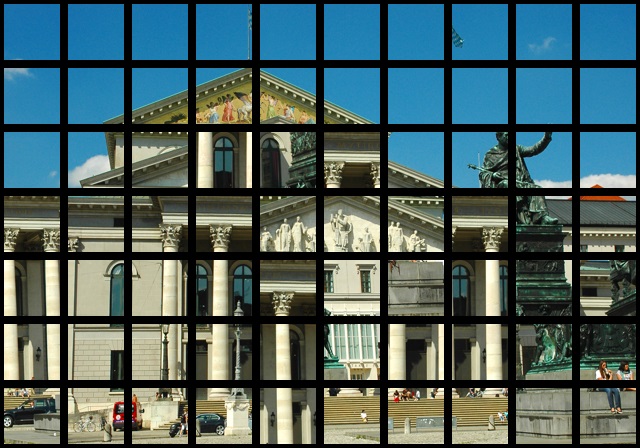}\\
\hline

150 pieces& & & \\
Erosion = $7\%$ &\includegraphics[width=0.24\textwidth]{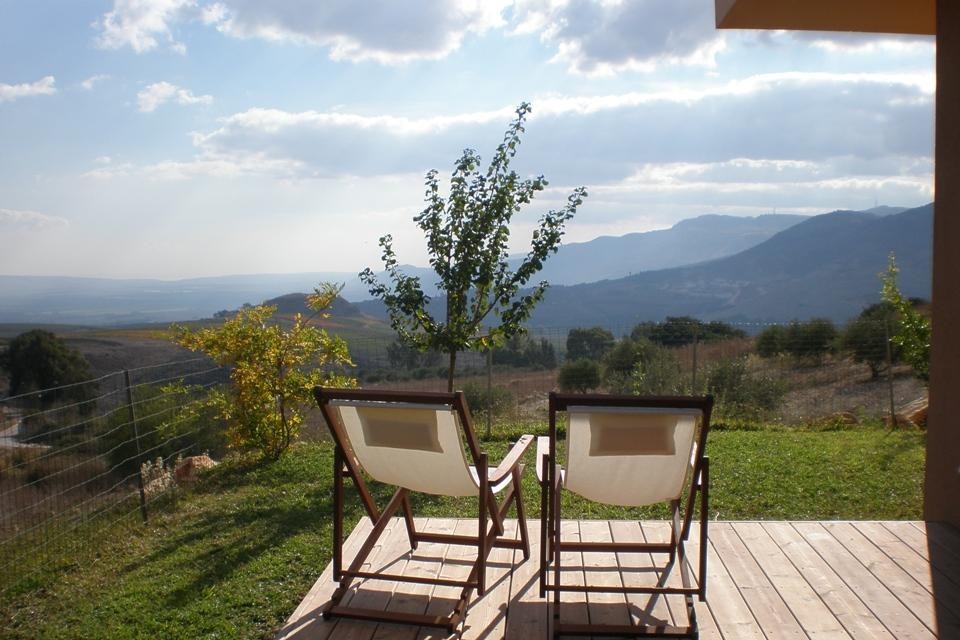}&
\includegraphics[width=0.24\textwidth]{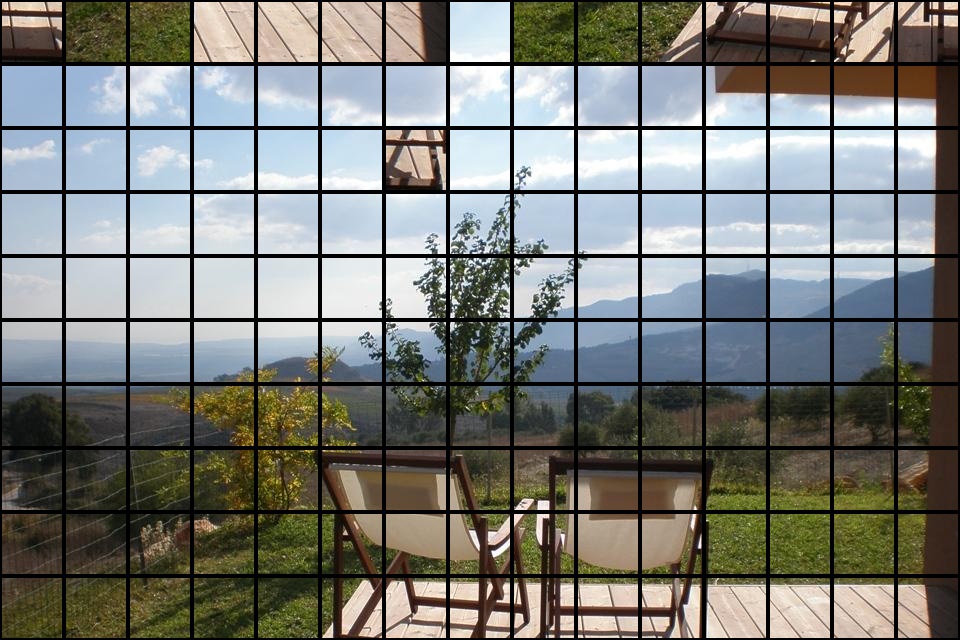}&
\includegraphics[width=0.24\textwidth]{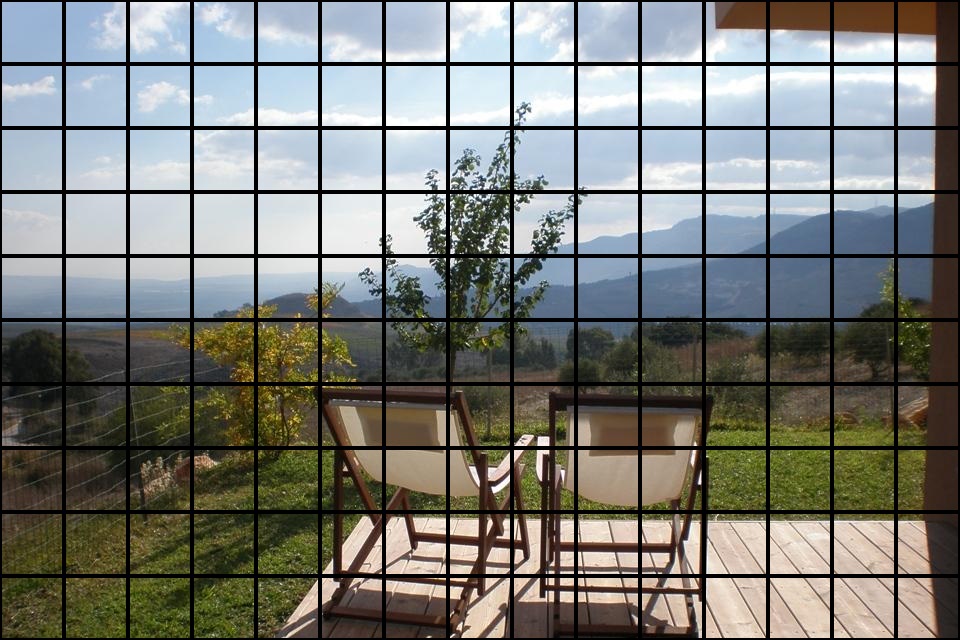}\\
\hline
Erosion = $14\%$ & &
\includegraphics[width=0.24\textwidth]{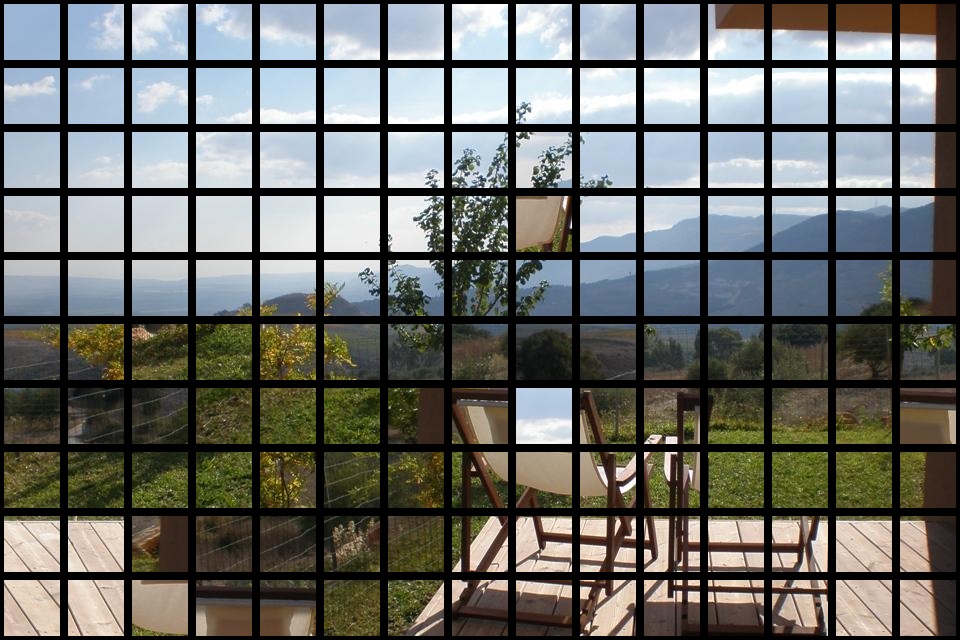}&
\includegraphics[width=0.24\textwidth]{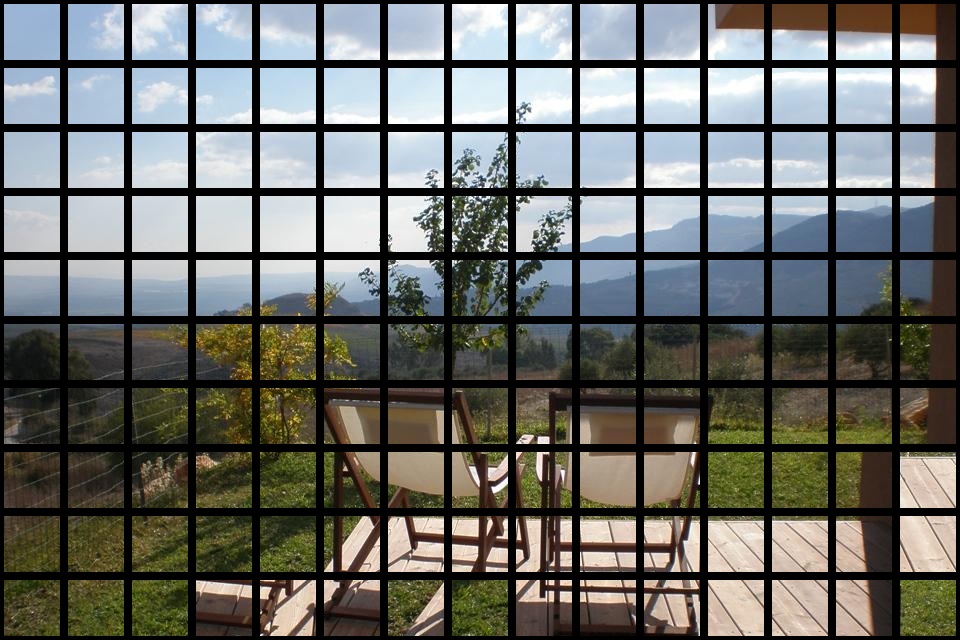}\\
\hline
\end{tabular}
\caption{
{\bf Qualitative results (cont').}
The larger the gaps, the worse the results.
However, the negative effects are more dramatic in~\cite{genady}'s results compared to ours.
}

\label{fig:qualitative_gap_8_continued}
\end{figure*}

Figure~\ref{fig:qualitative_gap_4} shows puzzle reconstruction results where the extent of the erosion in the gap is $7\%$ of the piece size.
It can be seen that our overall result is very good---almost identical in the eyes of a human observer.
Nevertheless, there are misplaced smooth sky pieces as well as sand pieces.
However, \cite{genady}'s mistakes are much more blatant; completely messing up the order of sand, people, water and sky.

Figures~\ref{fig:qualitative_gap_8}-\ref{fig:qualitative_gap_8_continued} illustrate the influence of larger gaps on the reconstruction, showing puzzles where the erosion extents are $7\%$ \& $14\%$ of  patch size.
As expected, the results get worse with larger erosion. 
Again, the negative effects of increasing the gap size are much more dramatic in~\cite{genady}.

Tables~\ref{table:scores_gap_4} and~\ref{table:scores_gap_8} report the quantitative results, averaged per dataset, using the three common measures~\cite{5probabilistic_jigsaw}. 
The average {\em neighbor measure} is considered the most important measure.
It computes the fraction of the correct pairwise adjacencies. 
The {\em direct measure} considers the fraction of the pieces that are in their correct absolute position. 
This measure is considered to be less meaningful due to its inability to cope with slightly shifted puzzles. 
Finally, the {\em perfect} columns indicate the number of puzzles for which the algorithms produced perfect reconstructions.
 Table~\ref{table:scores_gap_4} shows that our method performs much better than~\cite{genady} in all $3$ datasets, and in all $3$ measures. 
 In Table~\ref{table:scores_gap_8} results of both methods drop (compared to Table~\ref{table:scores_gap_4}) due to the increased difficulty.
 Our method consistently outperforms~\cite{genady}'s.
\begin{table}[t]
\begin{tabular}{ |p{1.6cm}||p{0.8cm}|p{0.8cm}||p{0.8cm}|p{0.8cm}||p{0.8cm}|p{0.8cm}| }
\hline
& \multicolumn{2}{|c||}{Neighbor} & \multicolumn{2}{|c||}{Direct} & \multicolumn{2}{|c|}{Perfect}\\
\hline
\# of pieces & \cite{genady} & Our & \cite{genady} & Our & \cite{genady} & Our\\
\hline
70 pieces   & 68.4\% & 84.6\% & 42.9\% & 86\% & 1 & 4\\
\hline
88 pieces   & 66.2\% & 76.9\% & 43.4\% & 70.7\% & 3 & 7\\
\hline
150 pieces  & 65.9\% & 76.3\% & 42.2\% & 66.7\% & 0 & 2\\
\hline
\end{tabular}
\caption{{\bf Quantitative evaluation (erosion extent of $7\%$).}
Our method outperforms~\cite{genady}'s in all $3$ common measures.}
\label{table:scores_gap_4}
\end{table}

\begin{table}[h]
\begin{tabular}{ |p{1.6cm}||p{0.8cm}|p{0.8cm}||p{0.8cm}|p{0.8cm}||p{0.8cm}|p{0.8cm}| }
\hline
& \multicolumn{2}{|c||}{Neighbor} & \multicolumn{2}{|c||}{Direct} & \multicolumn{2}{|c|}{Perfect}\\
\hline
\# of pieces & \cite{genady} & Our & \cite{genady} & Our & \cite{genady} & Our\\
\hline
70 pieces   & 41.4\%  & 57.1\% & 12.1\% & 50.5\% & 0 & 1\\
\hline
88 pieces   & 38.5\% & 51.1\% & 12.2\% & 35.2\% & 0 & 0\\
\hline
150 pieces  & 39.7\% & 51.3\% & 11.1\% & 35.4\% & 0 & 0 \\
\hline
\end{tabular}
\caption{{\bf Quantitative evaluation (erosion extent of $14\%$).} 
The results of both methods drop with larger gaps.
However, our method still outperforms~\cite{genady}'s.}
\label{table:scores_gap_8}
\end{table}

\noindent
{\bf Running Time.} Our tests are run using an {\em NVIDIA GeForce GTX 1080 Ti} GPU. 
It takes approximately $62$ seconds to solve a $70$ piece puzzle, $98$ seconds for $88$ pieces, and $286$ seconds for 150 pieces.
This time includes computing the pairwise dissimilarity between each pair
of puzzle pieces.
\\

\begin{figure}
    \centering
    \begin{tabular}{cc}
    \includegraphics[width=0.24\textwidth]{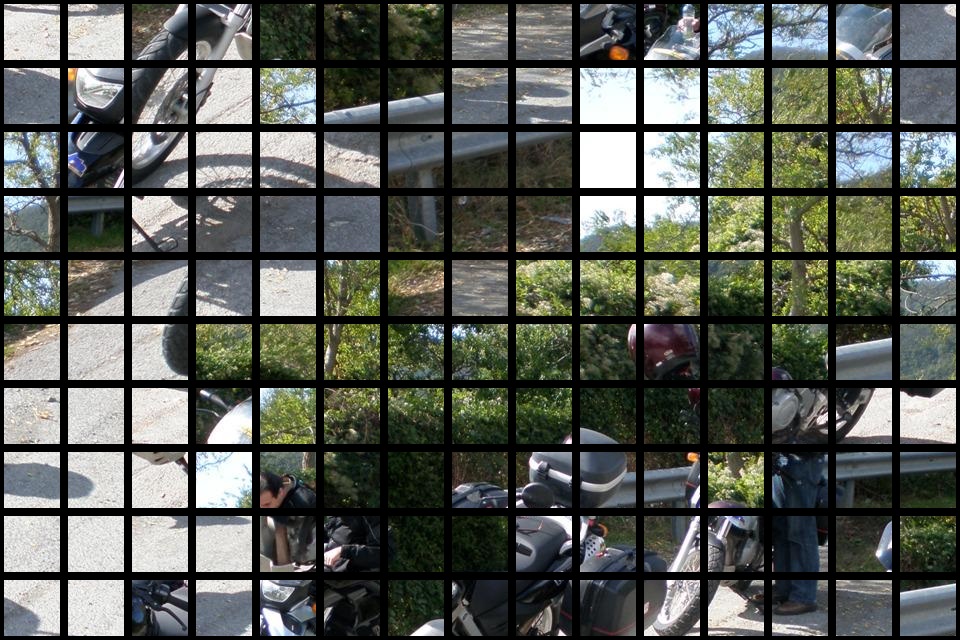}&
 \includegraphics[width=0.24\textwidth]{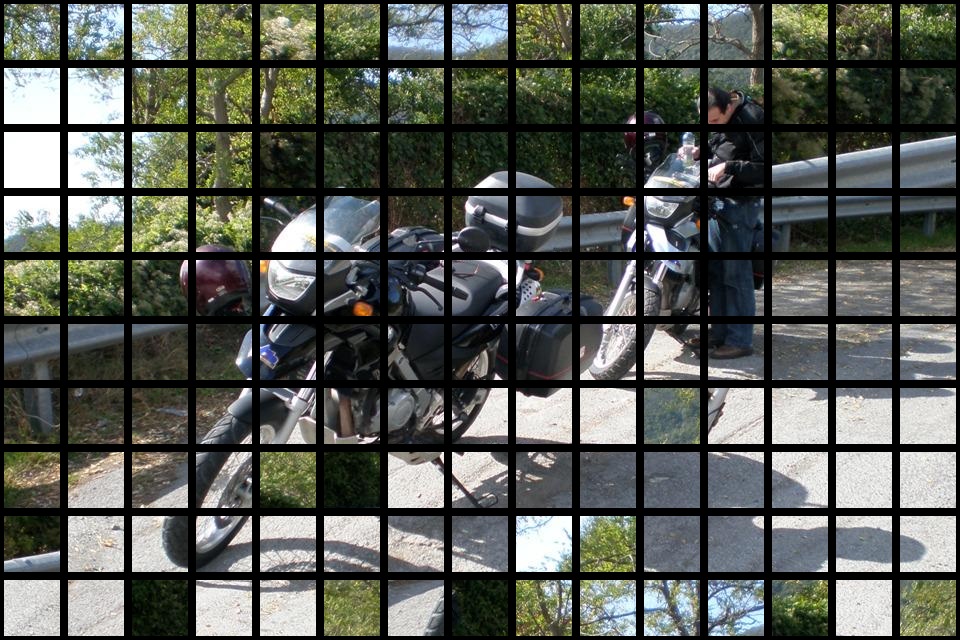}\\
 (a) \cite{genady}'s result & (b) Our result
 \end{tabular}
    \caption{{\bf Limitation.}Pieces of bushes and sky are misplaced in the solution of a $150$-piece puzzle.}
    \label{fig:limitation}
\end{figure}

\noindent
{\bf Limitations.}
Figure~\ref{fig:limitation} illustrates a case in which the puzzle is improperly solved by our method, as multiple pieces of bushes and sky are sporadically misplaced below the motorcycle.
Our solution is still preferable.

\section{Conclusion}
\label{sec:conclusion}

This paper introduced a novel algorithm for square jigsaw puzzle assembly, handling the extremely difficult case of eroded boundaries. 
While previous methods for solving square jigsaw puzzles heavily relied on color differences along the boundaries, we showed for the first time how deep learning can be advantageous in computing compatibility between eroded puzzle pieces. 

The key idea was to focus the learner's attention on the inpainting attempts of the erosion gaps. 
We introduced a GAN-based inpainting method that provided better results for our task than general SOTA methods.
Then, the quality of the inpainting served for neighbor classification.
This core idea, that when information is missing we can generate it and draw conclusions based on the generation quality, may find diverse applications.

We showed that our integrative method outperforms the state-of-the-art square jigsaw puzzle solver over all commonly-used metrics. 
Even when mistakes were made, they were much less noticeable to the human eye in our solution.
\newpage
{\small
\bibliographystyle{acm}
\bibliography{mainbib}
}
\end{document}